\documentclass[pre,twocolumn,showpacs,showkeys,preprintnumbers,amsmath,amssymb]{revtex4}

\usepackage{graphicx}
\usepackage{dcolumn}
\usepackage{bm}
\usepackage[hypertex]{hyperref}

\begin{document}

\title{Phase transition in the two-dimensional dipolar Planar Rotator model}

\author{L.A.S. M\'ol}
\altaffiliation{Permanent address: Departamento de F\'isica, Universidade Federal de Vi\c{c}osa, 36570-000, Vi\c{c}osa, Minas Gerais, Brazil.}
\email{lucasmol@ufv.br}
\affiliation{ Departamento de F\'isica, Laborat\'orio de Simula\c{c}\~ao, ICEX, UFMG, Caixa Postal 702, 31270-901, Belo Horizonte, MG, Brazil }
\author{B.V. Costa}
\affiliation{ Departamento de F\'isica, Laborat\'orio de Simula\c{c}\~ao, ICEX, UFMG, Caixa Postal 702, 31270-901, Belo Horizonte, MG, Brazil }

\date{\today}

\begin{abstract}
In this work we have used extensive Monte Carlo simulations and finite size scaling
theory to study the phase transition in the dipolar Planar Rotator model (dPRM) , also
known as dipolar XY model.
The true long-range character of the dipolar interactions were taken into account
by using the Ewald summation technique.
Our results for the critical exponents does not fit those from known
universality classes. We observed that the specific heat is apparently non-divergent
and the critical exponents are $\nu=1.277(2)$, $\beta=0.2065(4)$ and $\gamma=2.218(5)$.
The critical temperature was found to be $T_c=1.201(1)$. Our results are clearly distinct from those of a recent
Renormalization Group study from Maier and Schwabl [PRB {\bf 70}, 134430 (2004)] and agrees with the
results from a previous study of the anisotropic Heisenberg model with dipolar interactions in a bilayer system
using a cut-off in the dipolar interactions [PRB {\bf 79}, 054404 (2009)].

\end{abstract}


\maketitle

\section{Introduction}
The Planar Rotator model (PRM) in two dimensions, also known as XY model, is known
to have a critical line in the low temperature region~\cite{wegner,berezinskii,kosterlitz}.
The PRM is described by the following Hamiltonian: $H=-J\sum_{<i,j>}
\vec{S}_i\cdot \vec{S}_j=-J\sum_{<i,j>}(S^x_iS^x_j+S^y_iS^y_j)$, where $\vec{S}_i$ is 
a two dimensional vector $(S^x_i,S^y_i)$ defined in the sites $i$ of a two-dimensional
lattice and $<i,j>$ means that the
summation is to be evaluated for nearest neighbors sites.
As a prototype the PRM is expected to describe the magnetic properties of ferromagnetic
thin films where the spins lie in the film plane.
Although very simple, this model presents
some unusual characteristics, as the absence of spontaneous magnetization
for any $T>0$, which is a consequence of the Mermin and Wagner theorem~\cite{mermin-wagner}.
Thus, the system can not have a phase transition of order-disorder type,
nevertheless, there is still a phase transition in the model characterized by a change
in the spin-spin correlation function behavior. It is observed an
algebraic decay of the correlation function below a characteristic temperature, $T_{BKT}$,
above which the decaying is exponential. Besides that, the
correlation length is expected to diverge exponentially as long as
$T_{BKT}$ is approached from above, i.e., $ \xi(r) \sim a \exp{(b/\sqrt{T-T_{BKT}})}$
for $T>T_{BKT}$, while it remains infinity for any $T<T_{BKT}$.
This transition is named Berezinskii-Kosterlitz-Thouless
($BKT$) phase transition~\cite{berezinskii,kosterlitz}.
Several works, analytical as well as numerical, dealing with the subject were published since the
seminal work of Berezinskii and Kosterlitz and Thouless~\cite{minnhagen,hikami,takeno,evertz,costa}.
Besides that, it is also observed that the specific heat
does not diverge, instead, it has a broad maximum at a temperature slightly higher than
$T_{BKT}$~\cite{cuccoli,gupta,olson,lima}.
There are two interpretations for the mechanism leading to this transition:
Berezinskii~\cite{berezinskii} and Kosterlitz and Thouless~\cite{kosterlitz}
assume that it is driven by a vortex-anti-vortex unbinding mechanism,
while Patrascioiu and Seiler~\cite{patrascioiu} were able to obtain the
critical temperature and predicted the existence of a phase transition in the Coulomb
gas in any dimension ($d>1$) by considering that the mechanism responsible for the transition
is a polymerization of domain walls. (As a matter of unification of language we use in this
paper the terminology $BKT$ for this kind of transition).

However, in order to achieve a deeper insight on the magnetic properties of
thin films, one has to include dipolar interactions between the magnetic moments
of the lattice. This inclusion changes the scenario drastically, as
discussed by Maleev~\cite{maleev}. The long-range dipolar interactions
stabilizes the magnetization at low temperatures in such a way that an
order-disorder phase transition is now expected to take place. In a recent paper, Maier and Schwabl~\cite{maier}
have analyzed the phase transition in the dipolar Planar Rotator model (dPRM)
by using renormalization group techniques. Their
results indicate that the dPRM belongs to a new universality class characterized
by an exponential behavior of the magnetization, susceptibility and correlation length.
Besides that, the specific heat was found to be non-divergent, like occurs in the $BKT$ phase transition.
In this work, we have used extensive Monte Carlo simulations to study the phase
transition in the dPRM. Our results clearly indicate that the transition is of order-disorder type and is
characterized by a non-divergent specific heat and unusual critical exponents.

\section{dipolar Planar Rotator model and Monte Carlo method}

The model we are interested in consists of a square lattice with dimension $L\times L$.
At each site we place a classical spin variable
$\vec{S}_i = (S^x_i , S^y_i)$ with $\vec{S}_i^2=1$. The interactions are defined by the following
Hamiltonian:
\begin{equation}
H=-J\sum_{<i,j>}\vec{S}_i\cdot\vec{S}_j+ D\sum_{i \neq j} \left [ \frac{\vec{S}_i \cdot
\vec{S}_j}{r_{ij}^3}-\frac{3 (\vec{S}_i \cdot
\vec{r}_{ij})(\vec{S}_j \cdot \vec{r}_{ij})}{r_{ij}^5} \right ].
\label{hamiltonian}
\end{equation}
Here, $J>0$ defines a ferromagnetic exchange constant and $D$ is the dipolar constant.
$\vec{r}_{ij}$ connects sites $i$
and $j$ while $<i,j>$ means that the first summation
is evaluated for nearest neighbors only. For the dipolar
interactions the summation is evaluated over all pairs $i \neq j$.
Periodic boundary conditions have been used in the film plane ($x$ and $y$ directions) while open boundary conditions
were applied in the $z$ direction.
Ewald summation techniques~\cite{weis,wang} have been used to take into account the
true long-range character of the dipolar interactions\footnote{The Ewald summation allows one
to evaluate the dipolar energy without cutoffs, and details about this method can be found in Refs.~\cite{weis,wang}.}.
In all simulations we have assumed $J=1$ and $D=0.1$ and for these values
only ferromagnetic configurations were found in the low temperature regime. In this work the energy
is measured in units of $JS^2$ and temperature in units of $JS^2/k_B$.

Our Monte Carlo procedure consists of a simple Metropolis algorithm~\cite{metropolis,livro_landau}
where one Monte Carlo step (MCS) consists of an attempt to
assign a new random direction to each spin in the lattice. To equilibrate the system
we have used $100\times L^2$ MCS which has been found to be sufficient
to reach equilibrium, even in the vicinity of the
transition. In our scheme, two sets of simulations have been performed.
In the first one, we preliminarily explored the thermodynamic behavior of the model
in order to estimate the position
of the maxima of the specific heat and susceptibilities and the crossings of the fourth
order Binder's cumulant. In this first approach we used lattice sizes
in the interval $20 \leq L \leq 50$. Once the possible transition temperature is
determined, we refined the results by using single and multiple histogram
methods~\cite{ferrenberg,ferrenberg2}. We produced the histograms
for each lattice size in the interval $20 \leq L \leq 120$ and they were
built at/close to the estimated critical temperatures corresponding to the maxima and/or crossing
points obtained in step 1.
Details of the histogram techniques can be found in Refs.~\cite{ferrenberg,ferrenberg2}.

\section{Thermodynamic quantities and finite size scaling theory}

We have devoted our efforts to determine a number of thermodynamical quantities,
namely, the
specific heat, magnetization, susceptibility, fourth order
Binder's cumulant and moments of magnetization as described below.
The specific heat is defined as:
\begin{equation}
c_v=\frac{\langle E^2 \rangle - \langle E \rangle ^2}{Nk_BT^2},
\end{equation}
where $E$ is the internal energy of the system (computed using
equation \ref{hamiltonian}) and $N=L^2$ is the lattice volume.
The magnetization is:
\begin{equation}
M=\frac{1}{N}\langle m \rangle
\end{equation}
where
\begin{equation}
m=  \sqrt{ \left (\sum_{i=1}^N S^x_i \right)^2 +
 \left (\sum_{i=1}^N S^y_i \right)^2 }.
\end{equation}
The susceptibility is defined by the magnetization fluctuations
as:
\begin{equation}
\chi_{xy}=\frac{\langle m^2 \rangle - \langle m \rangle^2}{Nk_BT}.
\end{equation}
The fourth order Binder's cumulant reads:
\begin{equation}
U_4=1-\frac{\langle m^4 \rangle}{3\langle m^2 \rangle^2}.
\end{equation}
In order to calculate the critical exponent $\nu$, we also define the following
moments of the magnetization~\cite{kunchen}:
\begin{subequations}
\label{vjs}
\begin{eqnarray}
V_1 &\equiv& 4[m^3]-3[m^4], \\
V_2 &\equiv& 2[m^2]-[m^4], \\
V_3 &\equiv& 3[m^2]-2[m^3], \\
V_4 &\equiv& (4[m]-[m^4])/3, \\
V_5 &\equiv& (3[m]-[m^3])/2, \\
V_6 &\equiv& 2[m]-[m^2],
\end{eqnarray}
\end{subequations}
where,
\begin{equation}
[m^n] \equiv \ln \left| \frac{\partial \langle m^n \rangle }{\partial T} \right |.
\end{equation}

In critical phenomena the thermodynamic quantities are expect to behave
in the vicinity of the phase transition as~\cite{stanley,privman,livro_landau}:
\begin{subequations}
\begin{eqnarray}
c_v\sim t^{-\alpha} \\
\chi\sim t^{-\gamma} \\
M\sim t^\beta \\
\xi \sim t^{-\nu}, \label{xi}
\end{eqnarray}
\end{subequations}
where $t=|T-T_c|/T_c$ is the reduced temperature, $M$ is the magnetization,
$\xi$ is the correlation length and $\alpha$, $\beta$, $\gamma$ and $\nu$
are critical exponents. Although the critical temperature depends on
the details of the system in consideration, it is observed that the critical
exponents are universal, depending only on a few fundamental
factors~\cite{stanley,privman,livro_landau}. The systems are thus divided in
a small number of universality classes. Systems belonging to the same
universality class share the same critical exponents. Critical exponents are observed to
depend only on the spatial dimensionality of the system, the symmetry and
dimensionality of the order parameter, and the range of the interactions
within the system.

In a finite system as those used in Monte Carlo simulations
the divergences in the thermodynamic quantities are replaced by smooth
functions. Finite size effects are therefore of great importance
in the analysis of the results of Monte Carlo simulations. The theory
of finite size scaling~\cite{privman,livro_landau} provides one way to
extract information concerning the thermodynamic limit properties from results
obtained in finite systems. The basic assumption of this theory is that
in the vicinity of the phase transition the finite size effects should
depend on the ratio between the linear dimension of the system and
the correlation length say, $L/\xi$ . According to such a theory, specific heat,
susceptibility and magnetization for a finite system, in the vicinity of the
phase transition, behave as:
\begin{subequations}
\label{fss}
\begin{eqnarray}
c_v \approx c_\infty(t)+L^{\alpha/\nu}\mathbf{\mathcal{C}}(tL^{1/\nu}), \\
\chi \approx L^{\gamma/\nu} \mathbf{ \mathcal{X}}(tL^{1/\nu}),\\
M \approx L^{-\beta/\nu}\mathbf{ \mathcal{M}}(tL^{1/\nu}),
\end{eqnarray}
\end{subequations}
where $\mathbf{ \mathcal{M}}$, $\mathbf{ \mathcal{X}}$ and $\mathbf{ \mathcal{C}}$
are proper derivatives of the free energy. At $T_c$ ($t=0$) these functions are
constants and the size dependence of specific heat, susceptibility and magnetization
follow a pure power law. The size dependence of the pseudo-critical temperature,
$T_c(L)$, is~\cite{privman,livro_landau}

\begin{equation}
\label{tcrit}
T_c(L)=T_c+wL^{-1/\nu},
\end{equation}
where $T_c$ is the critical temperature in the thermodynamic limit.
Using the size dependence of the magnetization, equation \ref{fss},
and the definition of the moments of the magnetization in equation \ref{vjs},
one can easily show that such functions behave as:
\begin{equation}
\label{fss_vjs}
V_j \approx (1/\nu) \ln L + \mathcal{V}_j(tL^{1/\nu}),
\end{equation}
for $j=1,2,...,6$. At $t=0$ the functions $ \mathcal{V}_j(tL^{1/\nu})$ are constants
and then the curves for all $V_j$ have the same slope~\cite{kunchen} providing
a very precise method to determine both the critical exponent $\nu$ and the critical
temperature.

Concerning the fourth order Binder's cumulant, it is expected that
its curves should cross at the same point $U^*=U(T=T_c)$ for large enough $L$.
Besides that, its size dependence is expected
to obeys~\cite{binder81}:
\begin{equation}
\label{eq_u4}
U_4 \approx \mathcal{U}_4(tL^{1/\nu}).
\end{equation}

\section{Results}

\begin{figure}
\begin{center}
\includegraphics[scale=0.25]{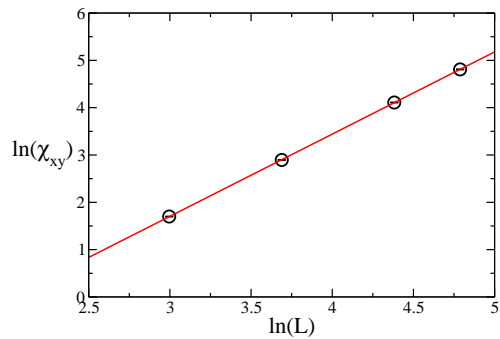}
\caption{\label{sus} Log-log plot of the maxima of susceptibility as a
function of the lattice size for $L=20,40,80$ and $120$. The error
bars are shown inside the symbols. The straight line is the best linear fit which gives
the exponent $\gamma/\nu=1.737(1)$.}
\end{center}
\end{figure}

\begin{figure}
\begin{center}
\includegraphics[scale=0.25]{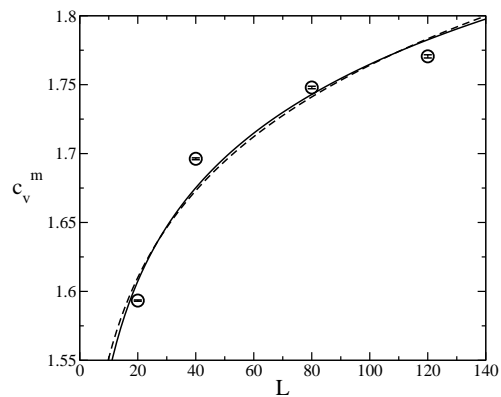}
\caption{\label{cv} Specific heat maxima as a function of the lattice size. The solid line 
is the best non-linear fit considering a logarithmic divergence and the dashed line
is the best fit considering a power law divergence. The error
bars are shown inside the symbols.}
\end{center}
\end{figure}

\begin{figure}
\begin{center}
\includegraphics[scale=0.25]{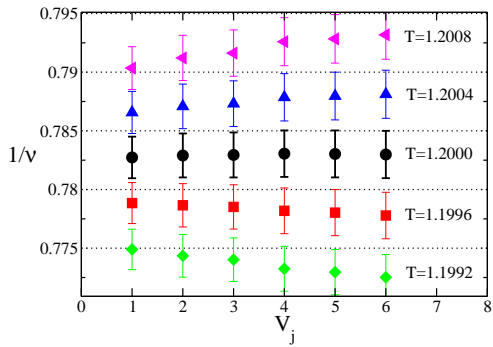}
\caption{\label{fvjs} Value of $1/\nu$ obtained by linear fits of
$V_j$ versus $\ln(L)$ for each value of $j$ at different temperatures.
Note that for $T = 1.2000$ the value of $1/\nu$ is almost the same for
all quantities.}
\end{center}
\end{figure}

\begin{figure}
\begin{center}
\includegraphics[scale=0.25]{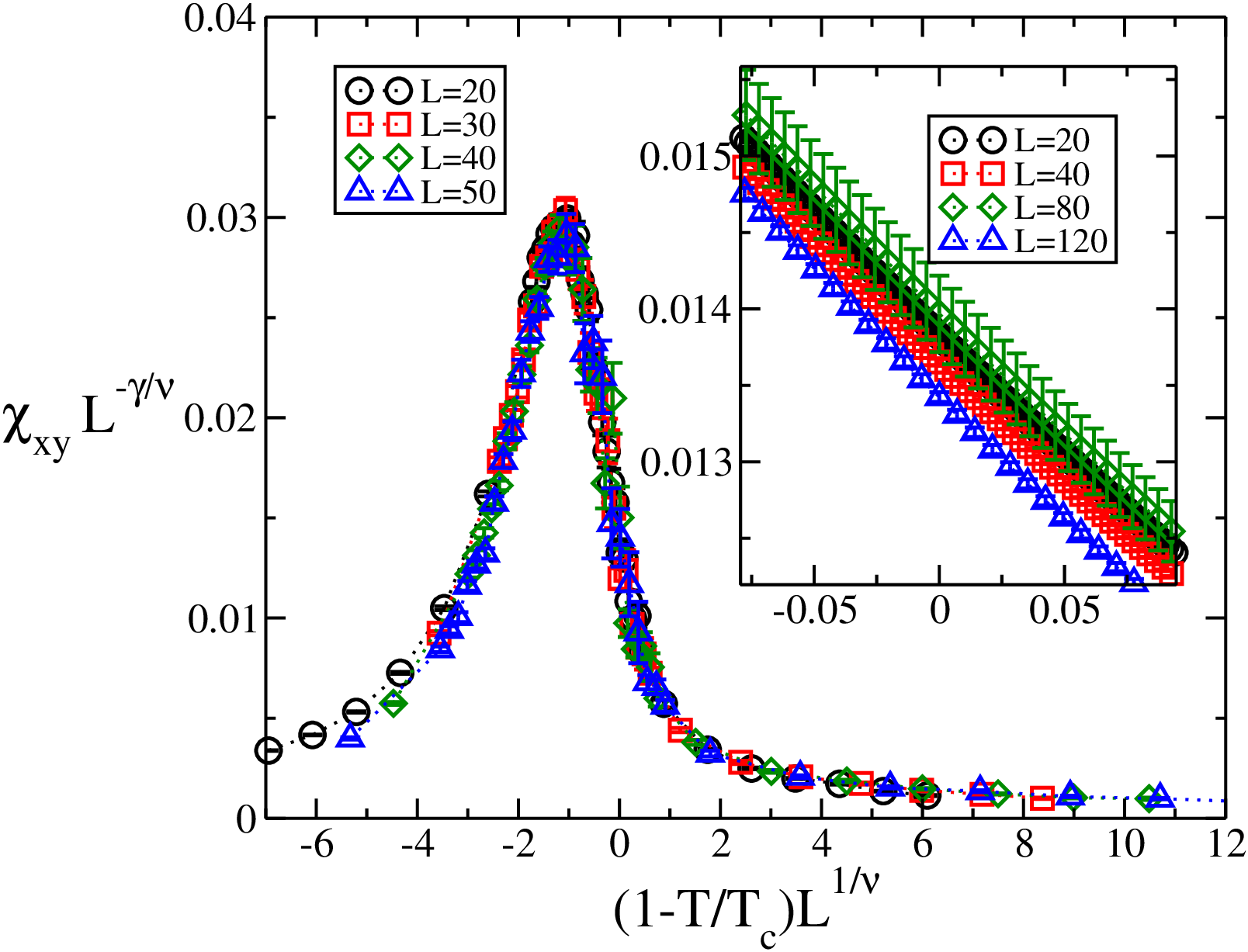}
\caption{\label{sc_sus} Scaling plot of susceptibility. According to
finite size scaling theory~\cite{privman,livro_landau} the susceptibility is expected
to behave as $\chi \approx L^{\gamma/\nu} \mathbf{ \mathcal{X}}(tL^{1/\nu})$. Note that the
curves for different lattice sizes collapses into a single curve.  In the outer plot the scaling
is done using results from conventional Monte Carlo simulations (step 1) for $L=20,30,40$ and $50$.
The inset shows the scaling for the histogram results (step $2$ in our simulations) for $L=20,40,80,120$.}
\end{center}
\end{figure}

\begin{figure}
\begin{center}
\includegraphics[scale=0.25]{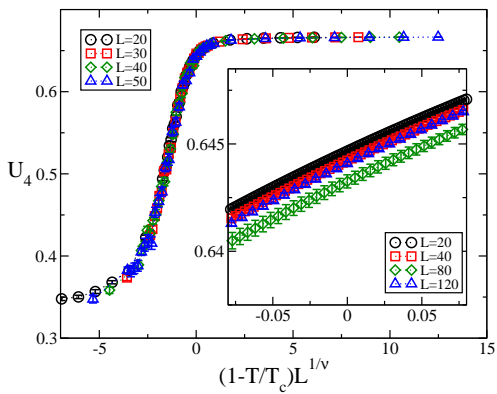}
\caption{\label{sc_u4} Scaling plot of the fourth order Binder's
cumulant. According to
finite size scaling theory~\cite{binder81} this quantity is expected
to behave as $U_4 \approx \mathcal{U}_4(tL^{1/\nu})$. Note that the
curves for different lattice sizes collapses into a single curve. In the outer plot the scaling
is done using results from conventional Monte Carlo simulations (step 1) for $L=20,30,40$ and $50$.
The inset shows the scaling for the histogram results (step $2$ in our simulations) for $L=20,40,80,120$.}
\end{center}
\end{figure}

In the following we show the results obtained by using the histogram method.
Each histogram consists of at least $3\times 10^7$ configurations.
In figure \ref{sus} we show a log-log plot of
the maxima of the susceptibility as a function of the lattice size for $L=20,40,80$
and $120$. The data are very well adjusted
by a straight line with slope $\gamma/\nu=1.737(1)$ exhibiting a power law
behavior. The specific heat maxima as a function
of the lattice size are shown in figure \ref{cv}. In this figure, the
solid line represents the best non-linear adjust of a logarithmic divergence
while the dashed one the best power law divergence adjust. It is clear that 
none of then can adjust our data satisfactorily. This result is similar to that obtained for the PRM
without dipolar interactions, and indicates a possible non-divergent specific heat.
In figure \ref{fvjs} we show the value of $1/\nu$ for some temperatures obtained by using
the moments of the magnetization defined in equation \ref{vjs} and \ref{fss_vjs}.
Using this method we get $T_c^{V_j}=1.1982(18)$ and $1/\nu=0.74(2)$.
%
%
%
%
%
%
%
%
%
%
%
\begin{figure}
\begin{center}
\includegraphics[scale=0.25]{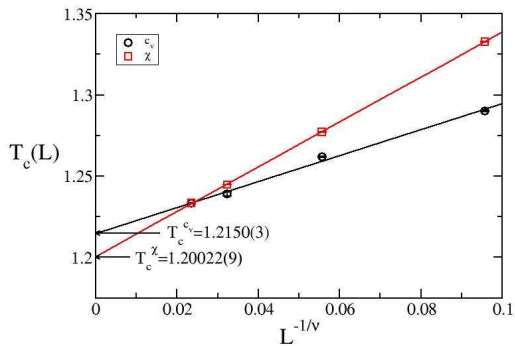}
\caption{\label{tc} $T_c(L)$ versus $L^{-1/\nu}$. From a linear adjust
we get $T_c^{\chi}=1.20022(9)$ and $T_c^{c_v}=1.2150(3)$.}
\end{center}
\end{figure}
%
%
%
\begin{figure}
\begin{center}
\includegraphics[scale=0.25]{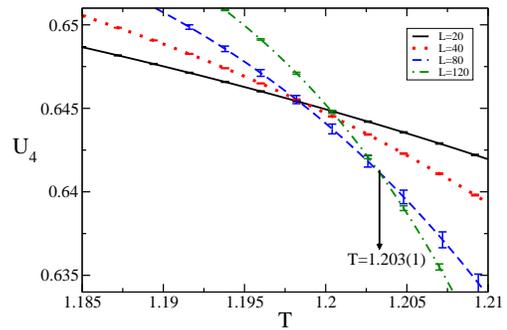}
\caption{\label{u4} Fourth order Binder's cumulant. The critical temperature
was estimated by the crossing point of the largest lattice sizes as being $T_c^{U_4}=1.203(1)$.
Only a few error bars are shown for clarity.}
\end{center}
\end{figure}
%
With the value of $1/\nu$ we may estimate the critical temperature
using the finite size scaling properties of the maxima of the susceptibility and
specific heat, see equation \ref{tcrit}. In figure \ref{tc} we show a plot of $T_c(L)$ as a function of
$L^{-1/\nu}$.
We obtain $T_c^{\chi}=1.20022(9)$ and $T_c^{c_v}=1.2150(3)$.
Using the crossing points of the fourth order Binder's cumulant~\cite{binder81}, see figure \ref{u4},
we estimate the critical temperature $T_c^{U_4}=1.203(1)$. 
Our best value for the critical temperature is thus the mean value of the
previous estimates $T_c^{U_4}$, $T_c^{V_j}$ and $T_c^\chi$ discarding the value
obtained by finite size scaling of the specific heat, since
its behavior is apparently non-critical.
This procedure gives $T_c=1.201(1)$. Plotting $\ln(M_{xy})$ versus $\ln(L)$ at $T_c$
it is possible to obtain the exponent $\beta/\nu$.
From a linear adjust we get $\beta/\nu=0.1617(2)$. In order to verify the
validity of our results we show in figures \ref{sc_sus}, (\ref{sc_mag})
and (\ref{sc_u4}), the scaling plots of the susceptibility, magnetization
and fourth order Binder's cumulant according to their finite size scaling
functions, see equation \ref{fss}. Note that all figures show a very 
good collapse of the curves for different lattice sizes.
%
%
%
%
%
\begin{figure}
\begin{center}
\includegraphics[scale=0.25]{fig7.eps}
\caption{\label{sc_mag} Scaling plot of magnetization. According to
finite size scaling theory~\cite{privman,livro_landau} this quantity is expected
to behave as $m \approx L^{-\beta/\nu} \mathbf{ \mathcal{M}}(tL^{1/\nu})$. Note that the
curves for different lattice sizes collapses into a single curve.  In the outer plot the scaling
is done using results from conventional Monte Carlo simulations (step 1) for $L=20,30,40$ and $50$.
The inset shows the scaling for the histogram results (step $2$ in our simulations) for $L=20,40,80,120$.}
\end{center}
\end{figure}
%
%
%
%
\section{Discussion}

In this work we have studied the phase transition in the ferromagnetic
dipolar Planar Rotator model (dPRM). Our results indicate that the
phase transition in this model is of order-disorder type and is
characterized by the exponents $\nu=1.277(2)$, $\beta=0.2065(4)$ and
$\gamma=2.218(5)$ and by a non-divergent specific heat. Our results
also indicate that the system have long-range order at low temperatures.
This conclusion is based on the following facts: ($i$) the magnetization
for $T<T_c$ does not display a significant decrease as the lattice
size is augmented, as for example, has been found in the Rapini's work~\cite{rapini1} and as expected for
a BKT phase transition; ($ii$) our results are very well described
by a finite size scaling theory based on the existence of a low temperature phase
with long-range order and finite correlation length~\cite{privman,livro_landau}.
In a BKT phase transition there is no long-range
order in the low temperature phase as a consequence of the Mermin-Wagner
theorem~\cite{mermin-wagner}. Indeed, the results of Maleev~\cite{maleev} 
predict the existence of long-range order at low temperatures in the dPRM and our
results are consistent with this scenario.

As discussed earlier, recent results by
Maier and Schwabl~\cite{maier} have predicted that this system may belong to
a new universality class, characterized by the presence of long-range order
at low temperatures and by an exponential behavior of thermodynamic
quantities in the vicinity of the ``critical'' temperature. By an exponential
divergence we mean that the correlation length diverges as the ``critical'' temperature ($T_c$)
is approached as $\xi \propto \exp (b/\sqrt{(T-T_c)})$, similarly to the behavior
of the BKT phase transition, while the behavior of other thermodynamic 
quantities are given by powers of the correlation length. 
Nevertheless, our results for the dPRM are very well described by
power law divergences of thermodynamic quantities. As can be seen
in figures \ref{sc_sus}, \ref{sc_mag} and \ref{sc_u4}, we obtained a very good
collapse of the curves from different lattice sizes for the susceptibility,
magnetization and Binder's cumulant. These curves show that the critical
exponents obtained and the conventional finite size scaling
theory, that assumes a power law behavior of thermodynamic quantities,
describe the Monte Carlo data accurately indicating that the phase
transition in the dPRM is a conventional order-disorder phenomena
with unusual critical exponents. 
In order to definitely rule out the
possibility of this phase transition of being in the new universality class
proposed by Maier and Schwabl, we should make a comparison of our
Monte Carlo results, using a finite size scaling theory based in their
predictions, and the conventional finite size scaling theory used here.
Unfortunately, it is not clear in the literature how
to obtain a finite size scaling theory for exponential divergences. 
Using a simple replacement of the correlation length by the lattice
size, which should be the first choice, does not give a good collapse
of the curves, mainly because the determination of the critical temperature
is quite imprecise in this case and the collapse of the curves depend
appreciably on the value used for the critical temperature. In any case,
using values for the critical temperature close to the maxima of the susceptibility
we were not able to obtain even a reasonable collapse of the curves.

Once the possibility of this phase transition being in
the new universality class proposed by Maier and Schwabl is discarded, some
questions arise: ($i$) Why Renormalization Group results
do not agree with our Monte Carlo simulations? ($ii$) Is the occurrence of
the order-disorder transition due to the long-range character of dipolar interactions
or to some other property of this model? A definite answer to these questions
may take a very long time to be given by virtue of the non-trivial characteristics
presented by this model.
Nevertheless, this study gave us some insight 
about what is happening. The RG study of Maier and Schwabl~\cite{maier}
is based upon some approximations, for instance, the using of a continuous
version of dPRM, where the lattice character is lost.
Since the dipolar interactions have an intrinsic
anisotropy which depends in a complicated manner on the location of each
spin in the lattice, the lattice geometry could have an strong effect
in the system. The identification and discussion of the finer points of the RG study
of the dPRM that cause the discrepancy in the results is beyond the scope
of this paper.
Concerning the origin of the order-disorder transition
the question is even more complicated. The long-range order observed at low temperatures
is expected to occur only when full long-range interactions are present.
Nevertheless, in a recent study of the anisotropic Heisenberg model
in a bilayer system~\cite{mol} using a cut-off in the dipolar interactions
we found the same critical behavior. In fact, the found critical exponents
($\nu=1.22(9)$, $\gamma=2.1(2)$ and $\beta=0.18(5)$) agree inside the
errors with those found in this study ($\nu=1.277(2)$, $\beta=0.2065(4)$ and
$\gamma=2.218(5)$). This observation indicates that the anisotropic character of
dipolar interactions may be the main responsible by the observed critical phenomena.
Indeed, this observation is not new in the literature. As an example,
Fern\'andez and Alonso~\cite{fernandez} stated that ``Anisotropy has a deeper effect
on the ordering of systems of classical dipoles in 2D than the range of
dipolar interactions''. In this work the authors found that the inclusion
of a quadrupolar anisotropy changes drastically the phase transition
behavior of a system of classical dipoles. Apparently, in our system
the intrinsic anisotropy of dipolar interactions play an essential role
in the determination of the universality class of the dPRM.

The possible new universality class is not surprising.
In the theory of critical phenomena~\cite{privman,livro_landau}
it is expected that the critical exponents, and thus the universality classes, depend only on the spatial
dimensionality of the system, the symmetry and dimensionality
of the order parameter, and the range of the
interactions within the system, characteristics not shared by the dPRM and models
of well known universality classes. To the best of our knowledge this work
and that of Maier and Schwabl~\cite{maier} are the only ones devoted
to the investigation of the critical behavior of systems with long-range
dipolar interactions (which are intrinsically anisotropic) and exchange
interactions in two dimensions.

\begin{acknowledgments}
We would like to thank Prof. D.P. Landau for
helpful discussions and W. A. Moura-Melo for a careful reading
of the manuscript. Numerical
calculation was done on the Linux cluster at Laborat\'orio de
Simula\c{c}\~ao at Departamento de F\'isica - UFMG.
We are grateful to CNPq and Fapemig (Brazilian agencies) for
financial support.
\end{acknowledgments}

\bibliography{paper_revtex}

\end{document}